# A Scintillator Purification System for the Borexino Solar Neutrino Detector

J. Benziger[a], L. Cadonati[b,1], F. Calaprice[b], M. Chen[b,2], A. Corsi[c], F. Dalnoki-Veress[b], R. Fernholz[b,3], R. Ford[b,4], C. Galbiati[b], A. Goretti[b], E. Harding[b,5], Aldo Ianni[c], Andrea Ianni[b], S. Kidner[b], M. Leung[b], F. Loeser[b], K. McCarty[b], D. McKinsey[b,6], A. Nelson[b], A. Pocar[b,7], C. Salvo[c], D. Schimizzi[a,8], T. Shutt[b,9], A. Sonnenschein[b,10]

[a] Chemical Engineering Department, Princeton University, Princeton, NJ 08544
[b] Physics Department, Princeton University, Princeton, NJ 08544
[c] INFN, Laboratori Nazionale di Gran Sasso, Italy

## Abstract

Purification of the 278 tons of liquid scintillator and 889 tons of buffer shielding for the Borexino solar neutrino detector is performed with a system that combined distillation, water extraction, gas stripping and filtration. This paper describes the principles of operation, design, and construction of the purification system, and reviews the requirements and methods to achieve system cleanliness and leak-tightness.

* Corresponding author. Tel: +1-609-258-5416; fax +1-609-258-2496
	Email: **benziger@princeton.edu**
1 Now at University of Massachusetts, Amherst, MA USA
2 Now at Queens University, Kingston, ON Canada
3 Now at Kingley, MI USA
4 Now at SNOLab Sudbury, ON Canada
5 Now at Lockheed Martin Corp. Sunnyvale, CA USA
6 Now at Yale University, New Haven, CT USA
7 Now at Stanford University, Stanford, CA USA
8 Now at Genetech, So. San Francisco, CA USA
9 Now at Case Western Reserve University, Cleveland, OH USA
10 Now at University of Chicago, Chicago, IL USA

# I. INTRODUCTION

The Sun is an intense source of neutrinos that are emitted in nuclear reactions that provide its power. Because neutrinos interact weakly with matter as they travel from the core of the sun to the earth, they can be used to directly study the nuclear reactions in the central core of the sun. Solar neutrinos also provide a unique opportunity to study fundamental properties of the neutrino[1]. An important effect concerning neutrino mass that can be studied is the influence of the high matter density of the sun on neutrino oscillations, in particular, the expected transition from matter enhanced oscillations to vacuum oscillations at neutrino energies below ~2 MeV[2].

Experiments on solar neutrinos began in the early 1960's with the chlorine experiment, the first to detect solar neutrinos[3, 4]. The chlorine experiment, and the subsequent gallium experiments detected solar neutrinos by radio-chemical methods [5, 6]. The chlorine and gallium experiments detected radioactive $^{37}$Ar and $^{68}$Ge atoms produced by neutrino interactions on a target of several tons of material containing $^{37}$Cl and $^{68}$Ga, respectively. This method measures an integral response of the neutrinos with energies above a threshold and is not sensitive to a specific neutrino branch. The radio-chemical detectors have the significant advantage that the signal is largely insensitive to α, β and γ radiations from natural radioactivity. The major limitation of the radio-chemical experiments is that the reaction rate is so small that a time integrated response was required precluding detection of short term variations in the neutrino flux.

The Kamiokande experiment was the first to achieve real time detection of solar neutrinos[7-9]. Kamiokande, and the follow-up Super Kamiokande experiment, measured the flux of the high energy $^{8}$B neutrinos by detecting the Čerenkov radiation produced by electrons scattered by neutrinos. The $^{8}$B neutrinos comprise a very small fraction of the neutrinos emitted by the sun. However, owing to the high energy of $^{8}$B neutrinos the Čerenkov light yield is sufficient to detect neutrino events above ~ 5 MeV. Additionally, background from natural radioactivity is much reduced at energies > 5MeV.

The SNO experiment employed heavy water ($D_2O$) to measure the charged current reaction D(ν,e+)nn induced by high-energy $^{8}$B neutrinos[10, 11]. The positron was detected by Cerenkov radiation. SNO also measured the neutral current reaction D(ν,ν)pn by detecting the neutrons. The measurements of the charged and neutral current reactions demonstrated that the deficit of neutrinos detected in the early solar neutrino experiments was due to neutrino oscillations [2, 12, 13]. Backgrounds from natural radioactivity that could produce neutrons were carefully controlled in the SNO experiment.

The mono-energetic 0.862 MeV $^{7}$Be solar neutrino is of particular interest for understanding nuclear reactions in the sun and for understanding neutrino oscillations. Particles scattered by neutrinos with energies below 1 MeV do not produce enough Čerenkov light to measure the $^{7}$Be , pep and pp neutrino fluxes. The Borexino experiment employs a liquid scintillator for detection of these low energy neutrinos by elastic scattering of neutrinos on electrons. Ubiquitous backgrounds from α, β, and γ radiation due to natural radioactivity, most of which appear at energies less than ~2 MeV, pose severe challenges for reducing those backgrounds to permit the detection of the low energy part of the solar neutrino spectrum.

The control and removal of trace levels of naturally occurring radioactive isotopes is the defining requirement for the successful measurement of low energy solar neutrinos in real time. We report here details of a scintillator purification system that was developed for the Borexino experiment. The system was successfully used to achieve unprecedented low background levels in the liquid scintillator[14].

The concept of on-line purification was demonstrated with a 5 ton liquid scintillation counting test facility (CTF) at the Laboratori Nazionali del Gran Sasso (LNGS)[15, 16]. The CTF contained the liquid scintillator within a 2 m diameter sphere constructed of 0.5 mm thick nylon. The scintillator containment vessel was suspended within 1000 tons of ultrapure water shielding. Scintillation light was detected with 100 phototubes with light collectors mounted in the water viewing the scintillator vessel. The liquid scintillator tested in the CTF was the same as that employed in Borexino, a solution of 1.5 g/L of PPO (2,5 diphenyloxazole) in pseudocumene (1,2,4 trimethylbenzene, abbreviated as PC). This particular mixture was chosen based on relative simplicity for purification as well as having appropriate scintillation properties for Borexino. A description of the CTF and the purification system it employed for PC/PPO scintillator purification is published elsewhere[17-19].

More recently an alternative scintillator comprised of PXE (phenyl-o-xylylethane) and PTP (para-Terphenyl) was tested in the CTF[20]. This scintillator mixture was better density matched to water which might have permitted using a water buffer in Borexino. The PXE solvent has a much higher boiling point than PC (295ºC for PXE vs. 169ºC for PC) which made it more difficult to purify PXE with distillation. Adsorption of impurities in the PXE scintillator onto high purity silica gel column was shown to be successful with the CTF for batch processing[21, 22].

## II. THE BOREXINO SOLAR NEUTRINO DETECTOR

The design of Borexino is based on the principle of graded shielding with the scintillator at the center of a set of concentric shells of increasing radiopurity (see Fig. 1). The 278-ton scintillator is contained within a thin (125 μm) nylon Inner Vessel (IV) with a radius of 4.25 m[23]. Positions of scintillation events are reconstructed based on photon arrival times at the PMTs. Software is then used to choose events in a fiducial mass of 100 tons within the central core of the inner vessel. A second nylon outer vessel (OV) with radius 5.50 m contains a passive shield composed of pseudocumene and 5.0 g/L DMP (dimethylphthlate), a material that quenches the scintillation of PC to suppress the γ-emission from the photomultiplier tubes. The OV acts as a barrier against radon and other backgrounds originating from outside. A third vessel is a stainless steel sphere (SSS) with radius 6.85 m that contains the PC-DMP buffer fluid in the space between the OV and the SSS. The SSS also serves as a support structure for the PMTs. Finally, the entire detector is contained in a tank (radius 9 m, height 16.9 m) of ultra-clean water. The total liquid passive shielding of the central volume from external radiation (such as the rock) amounts to 5.5 m of water equivalent. Thin low-background ropes made of ultra-high density polyethylene hold the nylon vessels in place. The scintillation light is

December 24, 2007                                                                                                3

viewed by 2212 8" PMTs (ETL 9351) uniformly distributed on the inner surface of the SSS [24, 25]. All internal components of the detector were selected for low radioactivity. The predicted background due to external γ-rays in the fiducial volume and in the neutrino window (250–800 keV) is less than 0.5 counts/(day - 100 ton)[17, 26].

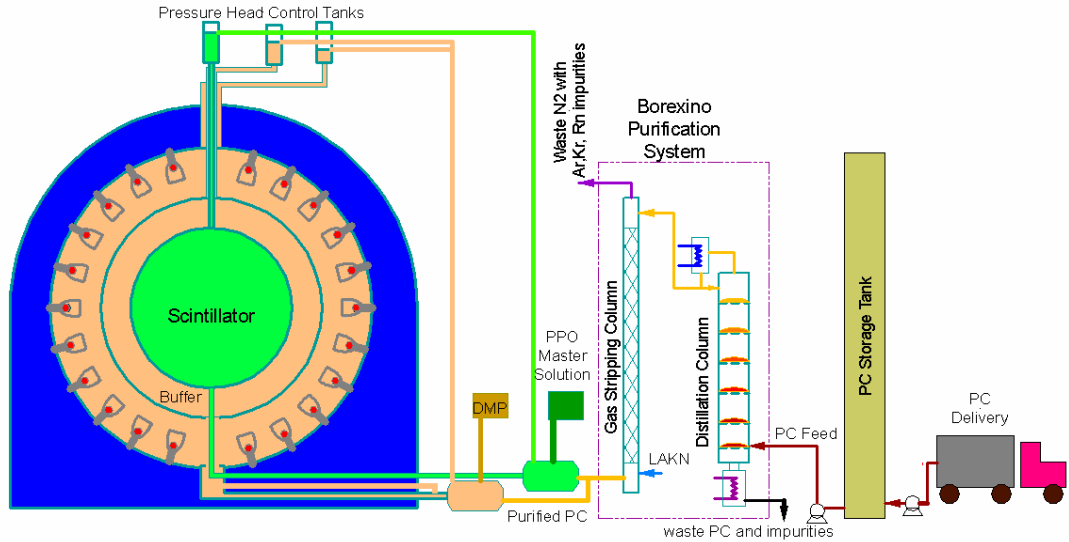

Figure 1. Schematic of the Borexino solar neutrino detector at Gran Sasso. The entire detector is located 1400 m underground. The detector is at the left. The outer tank is filled with water. Inside the water tank is a stainless steel sphere that supports the phototubes and contains two pseudocumene buffer regions; at the center is a pseudocumene/PPO scintillator mixture. The delivery of pseudocumene and purification of the scintillator and buffer is illustrated to the right.

## III. BOREXINO PURITY REQUIREMENTS

Neutrinos interact weakly with matter. The standard solar model predicts the flux of 862-keV $^7$Be solar neutrinos at the earth's surface be ~ 4.3 x $10^9$ /cm$^2$/s. That flux produces about 0.5 events/day per ton scintillator from the $^7$Be neutrino-electron scattering[27-29]. Detection of this neutrino flux requires radiopurity of the scintillator summarized in Table I; at the specified levels the desired decay rate for each impurity is less that 0.01 counts/day/ton scintillator within the spectral window of 250-800 keV. For $^{238}$U and $^{232}$Th the decay chains to $^{206}$Pb and $^{208}$Pb are assumed to be in equilibrium and all the progeny with appropriate energies have been included. The $^{14}$C/$^{12}$C ratio must be ≤ $10^{-18}$ to avoid the tail from the $^{14}$C β-emission with an end point of 156 keV from contributing to the neutrino window.

Table I identifies the primary source of the various impurities and the typical impurity level that would be anticipated without purification or special handling. The $^{14}$C level shown corresponds to carbon in biomass; scintillators derived from petroleum are expected to have much lower $^{14}$C levels as petroleum has been shielded from cosmic radiation. The other radioactive impurities enter in as contaminants from the air or dust. The strategy for reduction of the radioactive impurities is also indicated in Table I.

Concentration levels of the radioactive impurities that produce a decay rate of 1 count per day per 100 tons of scintillator ( 1 cpd/100-ton) are listed in the last column of Table 1.

To achieve the purity levels needed for Borexino required extraordinary cleanliness and leak-tightness of the entire purification system. Three specific requirements were applied to Borexino.

1. Vacuum leak tightness for total system leak rate < $10^{-5}$ mbar-L/s
2. Suspended particulate counts in liquids flowing through the system should be below Military Standard 1246C level 30.
3. Washing with detergent and chelating agents followed by de-ionized (DI) water rinsing of all surfaces to reduce surface deposits.

Vacuum leak tightness was required to eliminate influx of $^{39}$Ar, $^{85}$Kr and $^{222}$Rn from air. The leak specification was determined by the maximum tolerable background of 0.01 cpd/ton-scintillator; this level is sufficiently below the the $^7$Be neutrino signal of ~50 cpd/100 ton (a rate of ~$10^{-5}$ Bq/ton). The $^{85}$Kr is the most problematic of the radioactive noble gas background, it is in high concentration in air and has a long lifetime; $^{39}$Ar has 100 times lower activity and $^{222}$Rn will decay away within a reasonable time period. Many of the purification processes were operated at reduced pressure making it essential the system be leak tight to avoid the influx of $^{85}$Kr. $^{85}$Kr is found in air at levels of ~ 1 Bq/m$^3$-air; the $^{85}$Kr contained in 10 cm$^3$ of air would give a background comparable to the neutrino rate. The total leak rate for air during the filling operations can be estimated based on the tolerable level of $^{85}$Kr and the time to fill the detector. The purification system processes approximately 1 ton/hr. Hence, to achieve the design goal the total leak rate must be < $10^{-8}$ m$^3$-air/hr.

$$[\text{Air Leak Rate}] < \frac{\left(1\frac{\text{ton-scintillator}}{\text{hr}}\right)\left(\frac{0.1\ \text{cpd-}^{85}\text{Kr}}{100\ \text{ton-scintillator}}\right)}{\left(10^5 \frac{\text{cpd-}^{85}\text{Kr}}{\text{m}^3\text{-air}}\right)} < 10^{-8}\ \frac{\text{m}^3\text{-air}}{\text{hr}} \approx 3\times 10^{-6}\ \frac{cm^3 - air}{s}$$

The purification system and the filling stations had numerous valves and fittings. Assuming an approximate count of 30 such potential leaks from valves and fittings and a safety factor of ten, we required the leak rate to be less than $10^{-8}$ mbar-L/s on each valve and fitting in the purification system (1 mbar-L/s = 1 standard cm$^3$/s).

Particulates are suspended in the flowing fluids due to Brownian motion. These particles come from a variety of sources, including dust and shedding of submicron particles off the surfaces of the fluid handling and purification systems. The composition of the particulates is variable depending on the source. For design purposes we assume the particulates are dust and have a composition similar to the earth's crust with ~1 ppm of U and ~ 10 ppm Th[30]. Given that composition for suspended particles the maximum amount of dust that is acceptable to keep the $^{238}$U decay chain background to < 1cpd/100 ton-scintillator is ~$10^{-5}$ g-dust/ton-scintillator. Precision cleaning involving detergents, chelating agents and rinsing with filtered water can reduce the suspended particulate count to the tolerable level[31-33]. Design goals for system cleanliness were based on Military Spec 1246C[34]. This classification corresponds to the maximum size of suspended particles; level 50 refers to a maximum of 10 - 50 μm particles per liter of liquid; it is also assumed that the particle size distribution is given by a log-log distribution. The background from U and Th from suspended dust particles can be estimated for different levels of cleanliness by assuming the dust contains U and Th at the same concentration of the earth's crust (see values in Table 1), and the density of the suspended dust particles is 2 g/cm$^3$. Table 2 is a summary of the total mass concentration of particulates, and the expected background from U and Th decay chains. The last column in Table 2 is the background level from U and Th contamination if the scintillator goes through a final 0.05 μm filter that is 99.9% efficient at removing particulates >0.1 μm.

Table I
Radiopurity Requirements for Borexino

| Radioisotope | Source | Typical Level in scintillator without purification | Removal Strategy | Design Level (<1 cpd/100 ton) |
|---|---|---|---|---|
| $^{14}$C | Cosmic Ray activation of $^{14}$N | $^{14}$C/$^{12}$C~10$^{-12}$. Corresponds to equilibrium from cosmic radiation at earth's surface | Petroleum derivative (Old Carbon) | $^{14}$C/$^{12}$C~10$^{-18}$ |
| $^{7}$Be | Cosmic Ray activation of $^{12}$C | 2.7 x 10$^3$ cpd/ton. Corresponds to equilibrium for cosmic ray activation of $^{12}$C to $^{7}$Be at earth's surface | Distillation and Underground storage of scintillator | <0.01 cpd/ton |
| $^{222}$Rn | Air and Emanation from materials | 1.3 x 10$^7$ cpd/ton  Rn absorption into PC from air with $^{222}$Rn=10 Bq/m$^3$-air. | Nitrogen stripping, Rn decay | <0.01 cpd/ton |
| $^{210}$Bi | $^{210}$Pb decay | 2 x 10$^4$ cpd/ton.  Corresponds to $^{210}$Pb decay after exposing surface of the containment vessel to air with 10 Bq/m$^3$ $^{222}$Rn for 1 year. | Surface cleaning | |
| $^{210}$Po | $^{210}$Pb decay | 2 x 10$^4$ cpd/ton.  Corresponds to $^{210}$Pb decay after exposing surface of the containment vessel to air with 10 Bq/m$^3$ $^{222}$Rn for 1 year. | Surface cleaning | |
| $^{238}$U | Suspended Dust, Organometallics | 10$^4$ cpd/ton (for the entire decay chain)  <10$^{-12}$g-U/g-scintillator  1 g-dust suspended in 1 ton of scintillator with 10$^{-6}$g-U/g-dust. | Distillation, Filtration | <10$^{-17}$g-U/g-scintillator |
| $^{232}$Th | Suspended Dust, Organometallics | 10$^4$ cpd/ton  <10$^{-12}$g-Th/g-scintillator  1 g-dust suspended in 1 ton of scintillator with 10$^{-5}$g-Th/g-dust. | Distillation, Filtration | <10$^{-17}$g-Th/g-scintillator |
| $^{40}$K | Contaminant found in fluor | 2700 cpd/ ton  ~10$^{-9}$ g-K/g-scintillator  Scintillator with 1.5 g-PPO/L; PPO has 10$^{-6}$g-K/g-PPO. | Water extraction, filtration and distillation of fluor solution | <10$^{-14}$g-K/g-scintillator |
| $^{39}$Ar | Air | 200 cpd/ton  Ar absorption into PC from air with $^{39}$Ar= 13 mBq/m$^3$-air. | Nitrogen stripping, Leak tight system | <500 nBq/m$^3$ – N$_2$ |
| $^{85}$Kr | Air | 4.3 x 10$^4$ cpd/ton  Kr absorption into PC from air with $^{85}$Kr= 1 Bq/m$^3$-air | Nitrogen stripping, leak tight system | <100 nBq/m$^3$-N$_2$ |

To achieve the desired U background in Borexino requires the cleanliness level to be better then level 25 if there is not filtration. With filtration of the scintillator class 100 would be acceptable. In the preparation of the Borexino purification system cleaning was done to achieve better than level 50 cleanliness (typically level 30 was achieved) and we then relied on filtration to improve the backgrounds to the required level.

During the fabrication and assembly of the purification system, and during periods when the system was being modified, the internal surfaces of the system could be exposed to air containing $^{222}$Rn. The radon decay results in $^{210}$Pb deposited onto surfaces exposed to air; $^{210}$Pb accumulates and subsequently $^{210}$Bi and $^{210}$Po emanate from the surfaces as the $^{210}$Pb decays ($t_{1/2}$ ($^{210}$Pb) = 22 yr)[35, 36]. The $^{210}$Bi creates a broad background from β-emission with an endpoint at 1.16 MeV($t_{1/2}$ ($^{210}$Bi) = 5 d). $^{210}$Po decays with a 5.4 MeV α-emission ($t_{1/2}$ ($^{210}$Po) = 138 d) that produces a scintillation peak at in the neutrino window at ~425 keV (the energy for α-particles is reduced by ~11 in the PPO/PC scintillator[37, 38]). Precision cleaning reduces the amount of $^{210}$Pb adsorbed on the surfaces of the purification system[39]. In addition, all stainless steel surfaces were acid etched with nitric acid or a formic acid/citric acid mixture prior to precision cleaning to minimize the $^{210}$Pb contamination. When any modifications to the purification system were done the exposed parts were all acid etched (metallurgists refer to this as pickling) and then precision cleaned.

Table 2
Predicted Backgrounds from Particulates

| Cleanliness Level | Total Particulate Mass (g-particulates/g-liquid) | $^{238}$U Background from Particulates (cpd/ton) | $^{238}$U Background with filtration (cpd/ton) |
|---|---|---|---|
| 1 | 1.0 x 10$^{-14}$ | 1.0 x 10$^{-5}$ | 1 x 10$^{-8}$ |
| 5 | 1.5 x 10$^{-12}$ | 1.5 x 10$^{-3}$ | 2 x 10$^{-6}$ |
| 10 | 1.5 x 10$^{-11}$ | 1.5 x 10$^{-2}$ | 2 x 10$^{-5}$ |
| 25 | 3.1 x 10$^{-10}$ | 0.31 | 3 x 10$^{-4}$ |
| 50 | 3.6 x 10$^{-9}$ | 3.6 | 4 x 10$^{-3}$ |
| 100 | 3.9 x 10$^{-8}$ | 39 | 8 x 10$^{-2}$ |
| 200 | 8.2 x 10$^{-7}$ | 820 | 1 |
| 300 | 4.2 x 10$^{-6}$ | 4200 | 5 |

To minimize the deposition of $^{210}$Pb after cleaning the purification system was purged with nitrogen. Nitrogen was extensively used in both the CTF and Borexino as a blanketing gas in vessels and for stripping operations as well. Limits on the Ar and Kr contamination permissible for Borexino were established by assuming the scintillator would come to equilibrium with the nitrogen blanket during filling operations. The Ar and Kr activity in the atmosphere[40, 41] and the limits required for the nitrogen are listed in Table 3. The Heidelberg Borexino group tested nitrogen supplies from different producers and identified a number of sources with sufficiently low Ar and Kr for Borexino[42].

Table 3
Purity Requirements for Nitrogen

| Impurity | Natural activity in air (Bq/m$^3$-air) | Purity Requirement (Bq/m$^3$-nitrogen) | Purity Requirement (mole fraction in nitrogen) |
|---|---|---|---|
| $^{39}$Ar | ~13 x 10$^{-3}$ | 40 x 10$^{-9}$ | <3 x 10$^{-7}$ |
| $^{85}$Kr | ~1 | 10 x 10$^{-9}$ | < 10$^{-14}$ |

# IV. REVIEW OF PURIFICATION METHODS

The challenge for purification of low background liquid scintillators is that, except for the noble gases, the "molecular identity" of the impurities is unknown. The concentrations of these impurities in liquid scintillators are too small for chemical analysis. Because we don't know many of the specific molecular forms of trace impurities in liquid scintillators, the methodology applied to purification was to identify the most likely molecular form of the potential radioactive impurities, and choose purification methods most applicable to those.

The Borexino detector has two different fluids that have different compositions and purity requirements. There are 278 tons of scintillator comprised of 99.8 wt% pseudocumene and 0.2 wt% PPO. The two buffer regions have the same composition of 99.4 wt% PC and 0.6 wt% DMP (dimethyl phthalate) and have a total mass of 889 tons. The DMP acts as a quencher to fluorescence in the buffer regions. Since both the scintillator volume and the buffer volume are >99% PC the purity of PC has to meet the highest standards. The PC also has to be free of organic impurities that affect the optical clarity of the scintillator.

To minimize the storage requirements for PC it was decided to combine delivery, purification and detector filling into a continuous process. Continuous distillation followed by gas stripping was chosen for PC purification. The fluor, PPO, was obtained from Sigma-Aldrich was known to have K contamination greater than acceptable for Borexino[18, 19]. The PPO is purified as a concentrated master solution of 13 wt% PPO in PC in advance of the detector filling. The PPO master solution is purified by a combination of distillation, gas stripping and filtration. The DMP was obtained with sufficient purity from Sigma-Aldrich to be used as received.

Distillation, water extraction and gas stripping are equilibrium staged processes, that remove impurities based on compositional changes associated with phase changes. Distillation and gas stripping are based on differences in the equilibrium composition between liquid and vapor; water extraction is based on a difference in equilibrium concentration between an organic liquid and water.

A "theoretical" or equilibrium stage assumes that the two contacting phases achieve equilibrium and then pass on to the next stage. The composition of the two phases exiting an equilibrium stage is found by combining the mass balance with the phase equilibrium constraint. Phase contacting units can have physically distinct stages such as trays in a distillation column or they can have continuous contacting as in a packed column. The physical units do not correspond directly to equilibrium stages. Finite interphase transport rates result in the physical device achieving some fraction of the equilibrium separation. The reader is referred to standard texts and reference works for details about the correlations between operating parameters and stage-efficiency[43, 44]. In this section we outline the basic principles for the purification processes of the PC and the PPO master solution, identifying the key system design constraints and control parameters. These semi-quantitative models were employed to guide the design of the equipment and determine the system operating parameters to be used during the purification and detector filling. In Section 4 we provide a more detailed description of the process equipment.

The impurities identified as potentially most troublesome for a low background scintillation detector were:

(i) micron and submicron particulates (dust) containing K, U, and Th
(ii) $^{222}$Rn emanating from materials of construction
(iii) $^{85}$Kr and $^{39}$Ar from air leaks
(iv) $^{210}$Pb deposited on the surface of the scintillator containment vessel after exposure to $^{222}$Rn (this produced background of $^{210}$Po and $^{210}$Bi)
(v) $^{210}$Po from an unknown source that is significantly greater than expected from the equilibrium with $^{210}$Bi
(vi) $^{40}$K in the fluor

Radioactive impurities, such as $^{238}$U, $^{222}$Rn, $^{210}$Po and $^{85}$Kr associated with distinct molecular identities can, in principle, be separated from the scintillator. In contrast $^{14}$C is part of the molecular structure of the scintillator and cannot be removed from the scintillator. The $^{14}$C/$^{12}$C level should be $\leq 10^{-18}$ to avoid having a tail that spills over into the neutrino window. Pseudocumene, the principal component of the Borexino detector, is derived from petroleum which has been buried deep in the earth for millions of years so most of the $^{14}$C has decayed. A small amount of $^{14}$C remains in petroleum that is believed to be derived from α-particles from decay of $^{238}$U in the rock formation initiating the two step process $^{13}$C(α,n) → $^{14}$N(n,p) → $^{14}$C reaction. The $^{14}$C level may vary from different sources of petroleum feedstock from which the scintillator components were derived. Test results with the CTF confirmed that the total $^{14}$C/$^{12}$C level in the PPO/PC scintillator was ~2 x $10^{-18}$, acceptable for Borexino[45]. The $^{14}$C/$^{12}$C is six orders of magnitude lower than that found in biomass formed at the earth's surface!

In addition to removing radioactive impurities, purification also served the key function of removing chemical impurities that degraded the optical quality of the scintillator and buffer in the Borexino detector. Exposure of the scintillator to oxygen causes oxidation of pseudocumene and PPO; many of the partially oxidized products absorb light in the 300-450 nm region reducing the scintillation light that arrives at the phototubes. The optical attenuation length of freshly distilled PC at the peak emission of the scintillator is 7.4 m at 375 nm[19, 46]. If the PC is kept free of oxygen and contaminants it retains good optical clarity indefinitely. However, stainless steel acts as a catalyst, such that exposure to oxygen from the air and stainless steel at room temperature for 3 weeks can reduce the optical attenuation length to <2 m at 375 nm as shown in Figure 2[47].

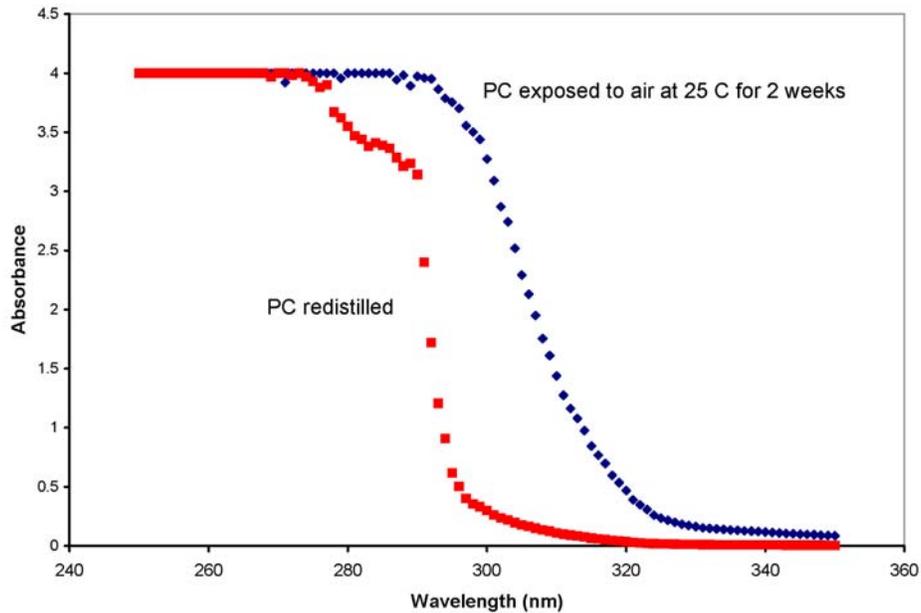

Figure 2. Optical attenuation length of PC. The plot compares PC that has been exposed to air and electropolished stainless steel for 2 week at 25 C to PC after 3-stage distillation. Both samples were thoroughly de-oxygenated with a nitrogen purge when obtaining the absorbance spectra.

## *IV.1. Distillation*

Distillation has been found to be the most effective process to improve the optical clarity of the scintillator; it is also highly effective at reducing several of the radioactive impurities in PC. The effectiveness of distillation at improving the optical clarity of PC is shown in Figure 2. After distillation the optical attenuation length at 320 nm increased from 1 m to > 10m. Distillation removes impurities that are less volatile than PC, but does not remove noble gas impurities. Distillation is followed by nitrogen stripping to remove Ar, Kr and Rn from the PC.

Other purification methods were tested for various liquid scintillators, including water extraction and adsorption[15, 16, 19-22]. Distillation generally separates the solvent and the fluor because they have different boiling points. Both water extraction and adsorption onto high purity adsorbents had the advantage of being able to process the scintillator mixture. Water extraction and adsorption were shown to be highly effective at removing radioactive metal impurities such as $^{238}$U, $^{232}$Th and $^{210}$Bi. Water extraction and adsorption showed only moderate success at reducing optical impurities. The effectiveness of water extraction and adsorption is due to polarity or charge of impurities that is common with metals or metal oxides. However, the optical impurities are often partially oxidized organic molecules with for which phase separation from the organic solvent of the scintillator is less favorable. The difference in purification efficacies of distillation, water extraction and adsorption motivated us to advocate distillation as the method of choice for the initial purification, while water extraction or adsorption might be preferred for re-purification.

Borexino employed a six stage distillation column operating at reduced pressure designed to process 1000 L/hr at a nominal operating pressure of 100 mbar and operating temperature of ~100 C; a simplified process flow diagram for continuous PC distillation and nitrogen stripping in Borexino is shown in Figure 3. PC distillation is a continuous operation with steady flows. A detailed description of the distillation hardware is presented in section VI.

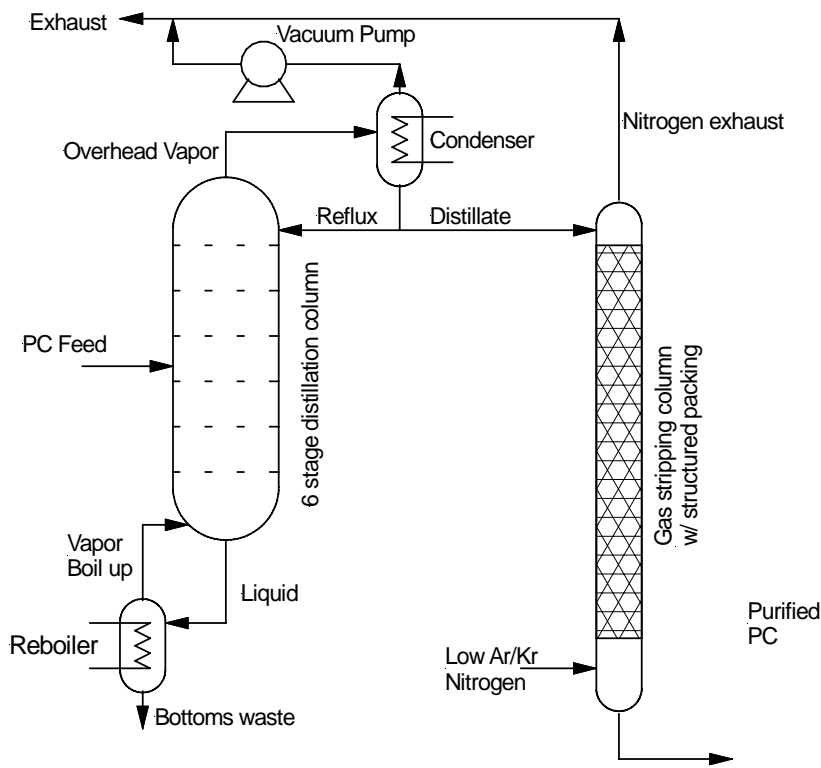

Figure 3. Process flow for PC purification in Borexino. The PC is distilled at reduced pressure in a six stage distillation column. The distillate product is fed to the top of a counter-current gas stripping column where high purity (low Ar/Kr ) nitrogen is used to strip $^{39}$Ar, $^{85}$Kr and $^{222}$Rn from the PC.

The three principal components of the distillation system are the column, the reboiler and the condenser. Liquid PC is fed to the column at tray 3 at a molar feed rate $F$. Liquid falls down the column by gravity through sieve trays which promote the contacting of vapors moving up through the column. Figure 4 is a photograph of a sieve tray during the assembly of the distillation column (assembly was done in a clean room with an Rn filter to keep from depositing $^{210}$Pb on the surfaces).

The liquid feed from tray 1 flows by gravity into the reboiler, which is an oil heated heat exchanger. In normal operation about 99% of the liquid entering the reboiler is vaporized and moves up through the distillation column. On each tray in the distillation column there is an exchange of material between the liquid and vapor in which the vapor is enriched with the more volatile components and the liquid is enriched in the less volatile components. The liquid and vapor flows must be kept within a limited operating range to make sure there is good contacting on the sieve trays. The size and number of the holes in the trays is based on nominal flow rates of vapor rising up the column and liquid fall down the column. If the flows are too high or too low, bypassing occurs and the poor contacting reduces the stage-efficiency.

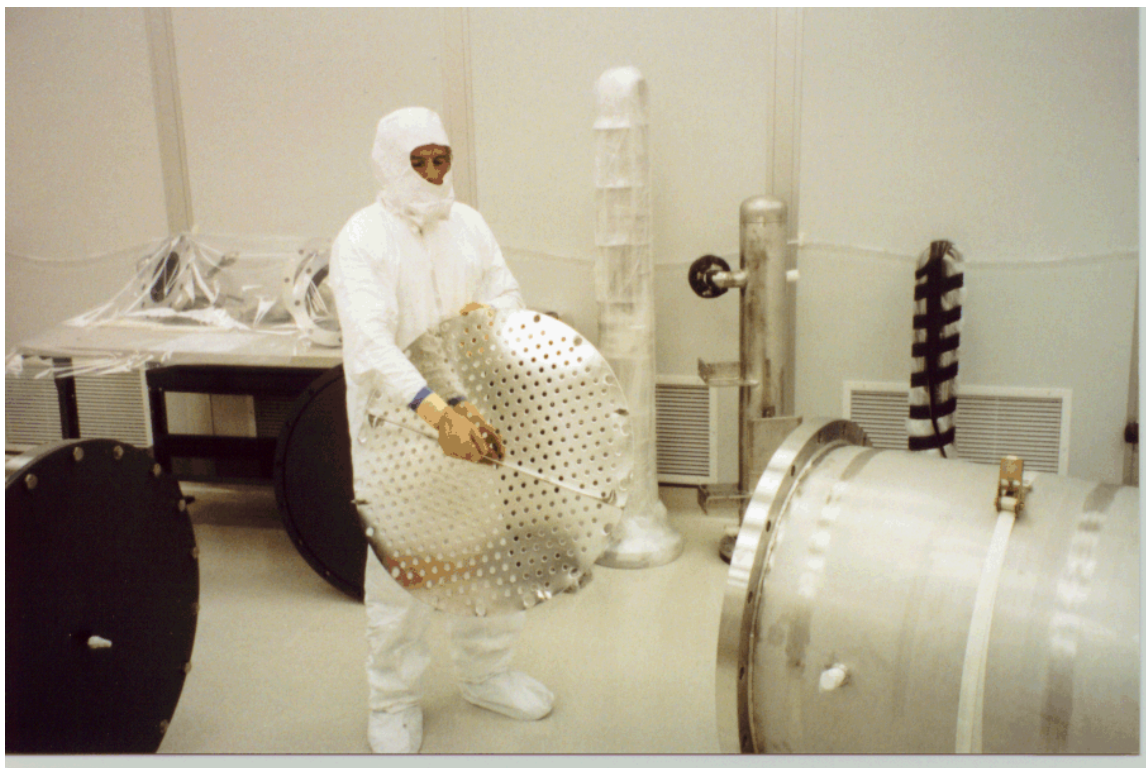

Figure 4. Installation of sieve trays into the Borexino distillation column. The column was assembled in a class 100 clean room with air filtered to reduce Rn to < 1 Bq/m$^3$. The trays lie horizontal. Liquid falls through the holes by gravity. Gas flow rising up through the holes holds the liquid up on the trays. The counter flow enhances mixing.

Vapor leaving the top of the distillation column (molar vapor flow rate $V$) enters a water cooled condenser where the PC vapor is condensed to liquid. The liquid from the condenser is split into two streams. The liquid distillate product, which will continue on to the gas stripping column, has a molar flow rate $D$. The remaining liquid from the condenser, molar flow rate $V-D$, is fed back to the top tray of the distillation column (tray 6) as reflux. The reflux ratio, $R$, is defined as the ratio of the quantity of reflux ($V-D$) to the quantity of distillate product ($D$). The reflux ratio is a key operating parameter that controls product purity; the higher the reflux ratio, the greater the product purity. At infinite reflux ratio the purity of the PC in the condenser is greatest, but there is no product being drawn off from the distillation column. At zero reflux the system throughput is maximized but the distillate product purity is at its minimum. The Borexino distillation column will be operated with a reflux ratio of 0.25, representing a compromise of product purity with reasonable throughput.

The effectiveness for distillation to remove impurities is based on the assumption that the impurities are less volatile (i.e. have a lower vapor pressure) than the PC. The effectiveness for separation depends on the relative volatility, $\alpha = P^o_{PC}/P^o_x$, which is the ratio of the vapor pressure of PC ($P^o_{PC}$) to the vapor pressure of the impurity ($P^o_x$) at the temperature of the distillation. The larger the relative volatility, the greater the purification achieved with distillation. In a multistage distillation column the purification factor per theoretical stage is approximately $\alpha \frac{R}{(1+R)}$. If there is no reflux (R=0), there is no liquid going down the column so there is only vapor moving up through the column and no purification takes place in the distillation column. At infinite reflux ratio the purification factor for each stage is equal to the relative volatility. For n theoretical stages above the feed the total purification factor may be approximated by equation 1, where $x_D$ and $x_F$ are the impurity concentrations in the distillate product and PC feed respectively.

$$\frac{x_F}{x_D} = \left[1 + \alpha \frac{R}{(1+R)}\right]^{n-1} \left[\frac{R+\alpha}{\alpha(R+1)}\right] \qquad [1]$$

(The interested reader is referred to references 40 and 41 for a detailed analysis of the mass balances for equilibrium stages in multistage distillation). The actual purification is less than the theoretical because of finite mass exchange between phases. At the design conditions of 1000 L/hr feed and reflux ratio 1, the 6 tray column was predicted to have 4 theoretical stages based on design correlations[48].

In addition to the reflux ratio, the purity of the PC is affected by the quantity of liquid discarded from the reboiler. If all the liquid were vaporized in the reboiler and no liquid discarded there would be no purification. In the Borexino purification system the feed comes in at tray 3 and the reboiler acts as an additional purification stage. The purity of the distillate product increases with the amount of bottoms, B, the liquid discarded from the reboiler. The minimum amount of bottoms that must be discarded may be approximated by assuming the reboiler is a theoretical stage with the composition of the vapor leaving the reboiler equal to the composition of the feed and the impurity is concentrated into the bottoms of the reboiler.

$$B > \frac{(R+1)}{\alpha} x_F F \qquad [2]$$

The impurities typically have much lower volatility than PC, typically $\alpha > 10^2$, so only a small fraction of the PC fed to the distillation column must be discarded. In the case of an impurity that has $\alpha \sim 1$ the required amount of bottoms that must be discarded can represent a substantial amount of the feed when the feed comes in at stage 1. The impurities could be further concentrated, reducing the amount of feed that must be discarded, by having a distillation column with the feed further up the column; multiple stages of distillation below the feed help to concentrate the less volatile impurities reducing the volume of bottoms that must be discarded. *Distillation stages above the* *feed reduce the impurity concentration relative to the feed, while distillation stages below the feed increase the impurity concentration relative to the feed.*

In summary the three key parameters that controlled the effectiveness of distillation were:

1 – Reflux ratio (increasing reflux ratio increases the purity of the PC product, but reduces the system throughput)

2 – Bottoms discard (increasing the amount of material discarded from the bottoms increases the product purity)

3 - System pressure (temperature). Decreasing the system pressure reduces the temperature of operation. The relative volatility increases with decreasing temperature which improves product purity. However, reduced pressure also reduces vapor flow rates in the distillation column, which reduces the efficiency of the interfacial mass transport.

To illustrate the effectiveness of distillation to purify PC, we consider the example of pseudocumene (vapor pressure 130 mbar at 102 C) contaminated with 0.1 mole % 2,4 dimethyl benzaldehyde (DMBA, vapor pressure 14 mbar at 103 C). DMBA is the initial oxidation product of PC formed by exposure to oxygen. DMBA has increased absorption of light around 350 nm, reducing the optical clarity of the scintillator, as seen in Figure 2. The relative volatility of PC to DMBA is ~ 9. The predicted stage efficiency is ~0.65 at the design conditions of F=1000 L/hr, a reflux ratio of 1 and bottoms discard of 1%. For four theoretical stages (6x0.65) the DMBA concentration in the distillate product is reduced by a factor of 90 compared to the feed. If the system were operated at a reduced flow of 800 L/h and 200 L/h reflux (reflux ratio 0.25) the predicted stage efficiency is reduced to ~0.5[49]. With three theoretical stages (6x0.5) and a reflux ratio of 0.25 the concentration of DMBA in the distillate is only reduced by a factor of 18. The increased reflux plays an important role improving the purification. The relative volatility of PC to known U or Th compounds is $>10^3$; with large relative volatility, it is possible to work with lower reflux ratios and still be very effective at removing heavy impurities from the PC with minimal bottoms discard required.

## *IV. 2. Gas Stripping*

The Borexino distillation system does not remove the impurities that are more volatile than PC; in particular, it will not remove argon, krypton or radon from PC. The noble gas impurities could be removed by a second distillation column where the PC was collected as the heavy component at the bottom of the column and the Ar, Kr and Rn were removed as the vapor out the top of the column. Rather than distillation the Ar, Kr and Rn impurities were removed by a separate gas stripping operation. Gas stripping is carried out with counter current flow of the liquid PC and gaseous nitrogen in a column containing structure packing (an example of the structured packing is seen in Figure 5). The process flow for gas stripping is shown in Figure 3. PC liquid flows down by gravity through the structured packing as a thin film with molar flow rate, $L_{PC}$. $N_2$ flows upward counter-current to the PC with molar flow rate, $V_N$. Equilibrium of the noble gases dissolved in the PC and the partial pressure of gas is described by Henry's law. If $y_\alpha$=mole fraction of $\alpha$ in the gas phase and $x_\alpha$=mole fraction of $\alpha$ in the liquid phase, then the system will evolve toward equilibrium where $x_\alpha H_\alpha = y_\alpha P$ where $H_\alpha$ is the Henry's law constant and P is the pressure of the system. When the gas starts as pure $N_2$ there will be a thermodynamics driving force for Rn, Kr and Ar dissolved in the PC to go into the gas phase. There will also be a driving force for $N_2$ to dissolve in the PC.

Henry's law constant for Rn, Ar and Kr in PC were estimated from values reported in the literature for benzene, toluene and xylene. Table 4 summarizes the Henry's law constants for Ar, Kr and Rn in PC at room temperature. The solubility of gases in liquids decreases slightly with increasing temperature (~ 1%/ºC), the temperature effect being greatest for the lightest gases, $N_2$ and Ar[50]. Because of reduced solubility gas stripping is more efficient at higher temperatures.

Table 4
Henry's Law Constant for Nobel Gases in Pseudocumene

| Gas | Henry's Law Constant in PC at 25ºC† (bar) | Henry's Law Constant for Water at 25ºC (bar) | Concentration Reduction for PC per stage at 30ºC, P=1 bar and $V_{N2}/L_{PC}=0.1$ | Equilibrium radioactivity in PC based on LAK nitrogen* at 1 bar (cpd/100-ton) |
|---|---|---|---|---|
| Ar | 1100[50] | 41,000[50] | 110 | <0.1 |
| Kr | 300[51] | 24,000[51] | 30 | <0.5 |
| Rn | 12[51] | 6,300[51] | 2.2 | N/A |

*LAK (Low Argon, Krypton) nitrogen contains 12 ppb Ar and 0.05 ppt Kr[42]

†Extrapolated from values for benzene, toluene and xylene[50, 51]

An upper limit approximation of the purification factor per equilibrium stage for gas stripping is given by equation 3 (for a more complete treatment of multistage gas stripping the reader is referred to references 40 and 41).

$$\frac{x_{\alpha,i+1}}{x_{\alpha,i}} = 1 + \frac{H_\alpha V_N}{P L_{PC}} \qquad [3]$$

By convention the feed enters at the top at stage n and flows down, the gas enters at the bottom at stage 1. The stages are numbered from the bottom up. The impurity concentration in the PC decreases with decreasing stage number in equation 3. The gas stripping packed column for Borexino was designed to operate at a liquid flow of 900 kg/hr (1 m$^3$/hr, 7.5 kmol/hr), a vapor flow of 20 kg-$N_2$/hr (0.71 kmol/hr) at 1 bar pressure. The purification factors predicted by Equation 3 at the nominal operating conditions are listed in Table 4. Ar and Kr are effectively removed from PC at atmospheric pressure. Rn stripping has a much lower efficiency for removal than either Ar or Kr. Radon has a high polarizability making it more soluble in the non-polar pseudocumene.

The ultimate purity of the PC from gas stripping is dictated by the thermodynamic limit given by Henry's law. As described in the previous section the nitrogen used for Ar, Kr and Rn stripping from the PC is carefully selected and shipped to have low levels of those impurities[42]. The concentrations of Ar, Kr and Rn in the $N_2$ used for stripping set a lower limit for the radiopurity that can be achieved by gas stripping. The last column in Table 4 lists the ultimate purity that could be achieved with $N_2$ stripping at 1 bar with the low Ar, Kr nitrogen.

It is evident from equation 3 that stripping at reduced pressure can improve the efficiency per stage of gas stripping, especially for Rn. However, the interfacial mass transport rate is substantially reduced in the absence of gas flow. Generally gas stripping is preferred to vacuum stripping because the gas flow improves the interfacial mass transfer between liquid and gas permitting higher stage efficiency. However, the ultimate purity with gas stripping is limited by the partial pressure of the impurities in the nitrogen used for the stripping. Reducing the pressure improves the theoretical purity for the same number of theoretical stages, but the equipment does not function as efficiently so it is harder to get to the theoretical purity limit.

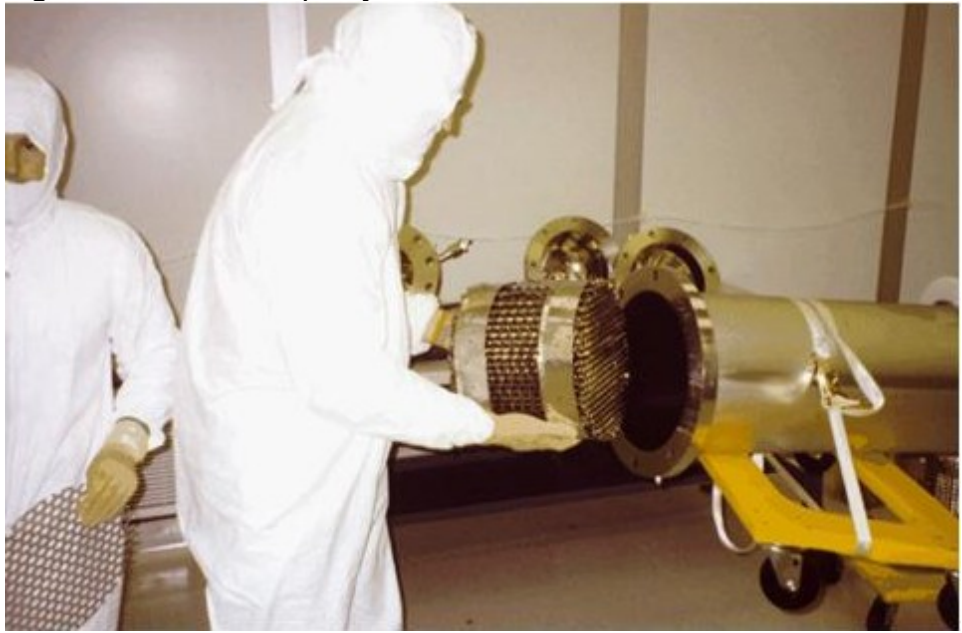

Figure 5. Structured packing being installed in the Borexino gas stripping column.

It is evident from equation 3 that stripping at reduced pressure can improve the efficiency per theoretical stage of gas stripping, especially for Rn. However, the interfacial mass transport rate is substantially reduced in the absence of gas flow. In a stripping column of fixed size there is an optimal pressure for gas stripping; reducing pressure increases the efficiency per theoretical stage, but reducing pressure also decreases the number of theoretical stages. Generally gas stripping is preferred to vacuum stripping because the gas flow improves the interfacial mass transfer between liquid and gas permitting higher stage efficiency.

During the filling of Borexino, reduced pressure stripping of the PC is planned. The optimal stripping operation is a balance between increased impurity reduction per stage and a reduced number of theoretical stages. Process simulations have been performed that suggest the optimal stripping pressure for Rn is between 100 and 200 mbar[52].

An alternative method of gas stripping using steam in place of nitrogen was tested for Borexino. Because water is not miscible with PC, the steam stripping was comparable to a vacuum operation, but the higher gas flow rates with the steam provided for more theoretical stages. The steam stripping was tested when it appeared that the typical nitrogen purity for Ar and Kr was not sufficient for Borexino. When a solution was found to control the Ar and Kr levels such that they were sufficient for Borexino the steam stripping was put on hold because it required more complex operations with steam generation and a condenser to recover the water (the water had to be recycled and purified because it was contaminated with PC).

Stripping at higher temperature and reduced pressure is also beneficial for reducing the nitrogen concentration in the PC after stripping. It was discovered in the laboratory that stripping PC at atmospheric pressure saturated the PC with nitrogen at the temperature of the stripping column. When the temperature of the PC rose so it exceeded the temperature in the stripping column, the nitrogen concentration in the PC was in excess of its solubility limit and nitrogen came out of solution and produced gas bubbles[15]. The gas bubbles accumulated and formed a gas dome in the detector. The buoyancy of the gas dome put extra stress onto the nylon vessels, which could endanger the detector vessel. By stripping at reduced pressure and elevated temperature the nitrogen concentration in the PC is kept below its solubility limit at the temperature and pressure of the detector.

The gas stripping column could also be used to prepare water for filling the Borexino vessels. There is concern that radon in the water could result in $^{210}$Pb being deposited onto the surface of the vessels, and this could eventually cause a problem with $^{210}$Bi and $^{210}$Po emanating from the surfaces giving rise to an undesirable background. With minor adjustments of the control system to accommodate the different physical properties of water (surface tension, viscosity, etc.) the gas stripping column could be used for water.

The number of theoretical stages for stripping of the PC with the Borexino system at the actual operating conditions was estimated to be between 2 and 3[48]. The purification factors per theoretical stage for Ar, Kr and Rn at a total pressure of 100 mbar and temperature of 30ºC are listed in Table 4. Also listed in Table 4 is the ultimate purity that can be achieved with the low Ar Kr nitrogen (LAKN) used for the stripping operation. It was estimated that the Ar and Kr levels approached their thermodynamic limit for stripping both water and PC. Radon reduction should approach the thermodynamic limit for water in the absence of radium or other radon sources dissolved in the water.

## IV.3. PPO Purification

Commercially available PPO was found to have K contamination at the ppm level, too high for Borexino[19]. Based on laboratory tests, distillation was the purification method of choice for the PPO. However, it is impractical to operate a continuous distillation of pure PPO because it solidifies below 70ºC. If all lines are not properly heat traced, PPO can condense in the transfer lines, causing the system to be blocked. Therefore, an alternative method was devised employing a concentrated master solution of 140 g-PPO in 1 L-PC.  This master solution is prepared in advance and then purified by flash distillation.  The process is illustrated in Figure 6.

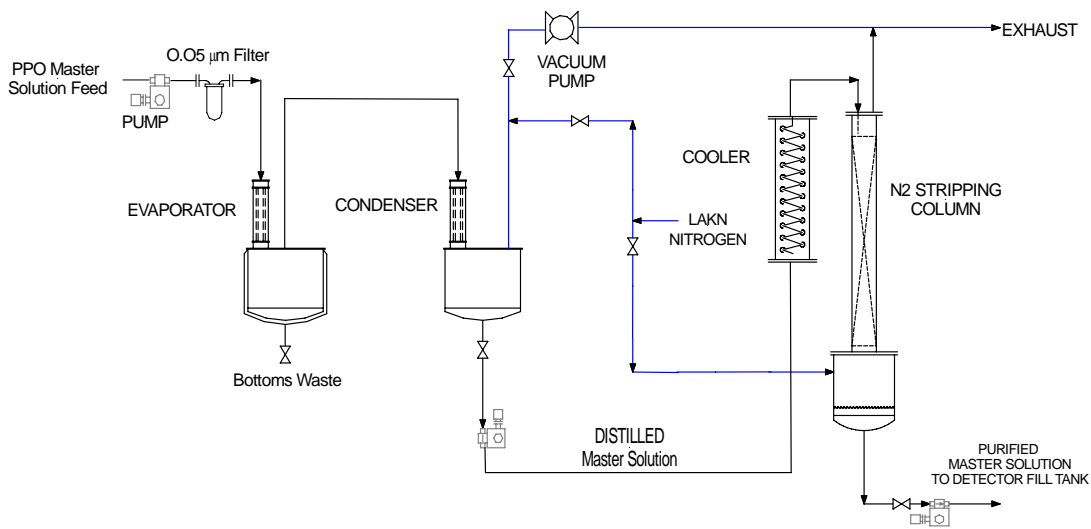

Figure 6.  Process flow for the semi-batch distillation of the PPO/PC master solution.

The relative volatility of PC to PPO is > 100 which presents a challenge in distilling the mixture.  To avoid excessive depletion of PPO from the master solution a semi-batch process is employed in which highly concentrated PPO (~99% PPO) is allowed to accumulate in the bottoms of a single stage distillation. The master solution is continuously fed to an evaporator operating at ~ 200ºC and 30 mbar.  Approximately 99% of the master solution is vaporized in an evaporator and the remainder is allowed to concentrate in the bottom of a heated vessel.  The vapor is passed to a condenser where the pressure is controlled by throttling nitrogen into a vacuum pump.  The liquid product from the condenser is then stripped by nitrogen in a column packed with stainless steel gauze.
      The bottoms from the evaporator are almost pure PPO; by allowing them to remain in the bottom of the evaporator the vapor in equilibrium with the liquid bottoms had a continually increasing PPO concentration.  Most of the PPO being fed to the evaporator is evaporated and passed on to the condenser.  There is a slow accumulation of PPO in the bottom of the evaporator along with low volatility impurities.  Periodically, the concentrated PPO that is accumulating in the bottom of the evaporator is discarded which also discards the heavy impurities.  PPO purification is planned to operate with a feed of 20 L/hr of 120 g/L PPO in PC.   1 L of PPO bottoms will be discarded approximately every 8 hours representing a loss of ~5.5% of the PPO.  The advantage of distilling the PPO/PC solution versus distilling PPO is that the master solution remains a liquid which avoids problems with solid condensation in the process lines.  Solid PPO can only form in the bottoms transfer line where the discarded PPO is discharged from the evaporator.  This short line can be heated to above 75ºC to keep the PPO in the liquid state and avoid plugging of the discharge line.

The distillation of the PPO master solution will only remove impurities that are less volatile than PPO. Gas stripping can remove impurities that are much more volatile than PC. PPO has a normal boiling point around 360ºC, so only impurities with higher boiling points than 400 C can be practically removed from the PPO master solution. Thus optical impurities like DMBA are not efficiently removed from the PPO/PC scintillator. It is therefore critical that the PPO master solution distillation be done with no air exposure. Vacuum tightness of the equipment is necessary not only from the viewpoint of contamination by Ar, Kr and Rn from the air, but also to avoid any oxygen exposure that can oxidize the PC and PPO because these impurities cannot be easily removed from the PPO master solution.

After distillation the PPO master solution will be stored in a vacuum tight container under a nitrogen atmosphere. The PPO master solution is metered to mix with freshly distilled PC during the filling of the Borexino detector. Details about the fluid handling operations are contained in a forthcoming publication[53].

# V. SCINTILLATOR RE-PURIFICATION

The purification system was built for on-line repurifying of the scintillator if unacceptably high levels of radioactive impurities are detected or if the optical quality of the scintillator unexpectedly degrades. Two strategies are accommodated by the purification system described above for repurification: water extraction followed by gas stripping, or distillation of the scintillator into PC and PPO, followed by repurification of the PPO and refilling the detector with on-line mixing. These two processes are illustrated in Figure 7, and summarized below. There are other proposals for scintillator re-purification, such as adsorption, for which construction of additional equipment would necessary.

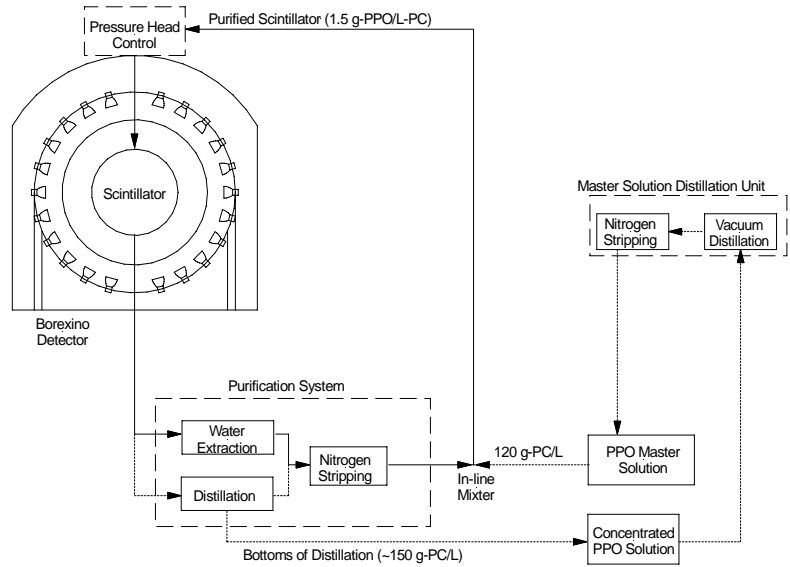

Figure 7. Process Flow Diagrams for Scintillator Repurification. The flow path for water extraction and nitrogen stripping of the scintillator is shown with a solid line. The flow path for distilling the scintillator into PC and PPO, coupled with PPO distillation and reconstituting the scintillator is shown with a dotted line.

The water extraction repurification process is similar to that applied in the CTF[15, 18, 19]. Scintillator is circulated from the detector to the purification plant where it is passed counter current to purified water in a packed column. The water extraction is effective at removing impurities that are polar or charged; such impurities have a greater affinity for water than the non-polar aromatic scintillator. Most of the inorganic radioactive impurities such as K, U, Th and Pb are expected to be present as particulates and ionic species which should be effectively removed by water extraction. After the water extraction the scintillator is stripped with nitrogen. The nitrogen stripping removes Ar, Kr and Rn impurities as well as removing the water dissolved in the scintillator. Water removal is essential to avoid liquid water phase separation from the scintillator if the temperature in the scintillator drops. Water phase separation is quite dramatic, causing micron sized water droplets to form and the scintillator appears cloudy. The water droplets cause light scattering in the scintillator which adversely affect the event position reconstruction.

Water extraction is not effective at removing many of the optical impurities that may be in the scintillator. Most of the optical impurities are partially oxidized PC or PPO which are more soluble in PC than water and will not be extracted. Water extraction is also not effective at removing some of the inorganic impurities. Results from bench top lab tests[39] and reports from the literature [54-56] suggest that some $^{210}$Po forms compounds that did not have a strong affinity for water and remained in the scintillator after water extraction. These impurities leave a $^{210}$Po background that is not in equilibrium with the $^{210}$Bi background.

The alternative repurification method is to split the scintillator back into its separate components, PC and PPO, and purify them separately as was done for the initial filling. The scintillator would be distilled in the purification plant separating the PC as the distillate product and concentrated PPO/PC solution as the bottoms product. The concentrated PPO/PC solution would be redistilled in the semi-batch process to recover a purified master solution. The PC distillate and the redistilled PPO/PC master solution would then be mixed in-line with the distilled PC and returned to the Borexino detector. It is planned to use a heat exchanger to heat the repurified scintillator to a temperature of ~20-25ºC, about 5-10ºC above the scintillator temperature in the detector. Because the repurified scintillator is at higher temperature it is slightly less dense than the scintillator in the detector and should remain partially stratified. Reducing the mixing of the scintillator in the detector during repurification should remove the impurities faster. (The ideal approach would be to displace the scintillator with water and repurify the scintillator in a batch type process. However, we do not have the capacity to store the entire PC from the detector and buffer regions so we cannot safely carry out a water displacement.)

# VI. DESIGN AND CONSTRUCTION OF THE PURIFICATION PLANT

During the period between April-June 1996, the design work for the purification plant was carried out in conjunction with Koch Modular Process Systems, Inc (Paramus, NJ). KMPS specializes in design and installation of purification systems for the petroleum, chemical and pharmaceutical industries. The engineers at KMPS provided expertise for equipment sizing, piping layout, and electrical and control systems based on the conceptual design and data from the Borexino collaboration. The purification plant was designed and built as a skid-mounted system for flexibility of installation. The process equipment was custom-fabricated stainless steel. KMPS oversaw the equipment procurement and system construction. Princeton reviewed and approved all materials and equipment selections and fabrication methods to ensure the system was leak tight and could be cleaned.

Equipment was designed and fabricated to a mix of standards from the electronics and pharmaceutical industries. The material of choice was 316L stainless steel, mechanically polished and electropolished, wherever possible. Teflon and quartz were also identified as alternative materials where stainless steel could not be used. Because of potential Ar, Kr and Rn contamination from the atmosphere the equipment was specified to be leak tight to a rate of $10^{-8}$ mbar-L/s. Large flanges (> 6") were all concentric double seal gaskets with valved connections to the space between the gaskets; the pump out ports were used for leak checking of the flanged seals. Wherever possible Helicoflex™ metal gasket seals were used, especially on any heated vessels. Elsewhere teflon encapsulated viton gaskets were employed. Process lines from ½" – 1" were all connected with VCR™ metal gasket fittings. Figure 8 shows an example of the double sealed flanges on one of the reboiler heat exchangers.

All instrument probes were either obtained with vacuum tight fittings or we fabricated custom fittings for a vacuum tight seal. Stainless steel diaphragm sealed valves were used throughout the system. Valves were obtained from a variety of vendors, primarily Swagelok and Carten-Banner. All the valves had specifications for leak-tightness to the outside of $< 10^{-8}$ mbar-L/s (the leak-tightness across the valve was not specified and as we discovered in several cases it can be greater than leaks from the outside). Nitrogen driven positive displacement diaphragm pumps constructed of teflon (White Knight) were used throughout. The pumps were sealed in enclosures blanketed with the exhausted nitrogen drive gas.

The plant was fabricated at Engineered Mechanical Systems (Elmwood Park, NJ) using the design of KMPS. The vessels and heat exchangers were manufactured by Tolan (Rockaway, NJ). Members of the Borexino collaboration from Princeton routinely visited Engineering Mechanical Systems during the construction phase to assure construction specifications were met. The large vessels and process equipment were all electropolished stainless steel. All the small tube welding was done by orbital welding. The larger welded vessels had the welds mechanically smoothed and then electropolished to a non-particulating finish. The surfaces were polished to a Ra 0.05 μm rating[57]. The structured packing employed in the gas stripping column and water extraction column were obtained from Koch-Glitsch. The packing material was produced from pressed sheets of 316 SS. It was not possible to electropolish the complex structured packing, but all the packing material was passivated and precision cleaned.

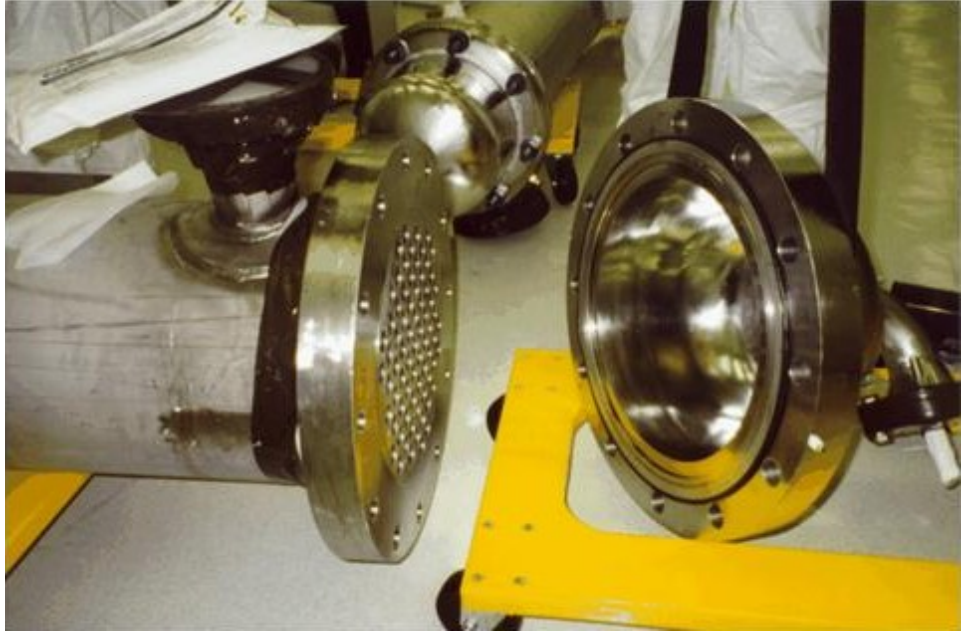

Figure 8. Close up view of the double gasketing system on flanges for leak tightness. The double gasketing system was employed throughout the purification skids. The outer gasket is a teflon encapsulated viton O-ring. The inner gasket is a metal helicoflex™ gasket. The mirror finish of the electropolished stainless steel used throughout the purification system is also evident.

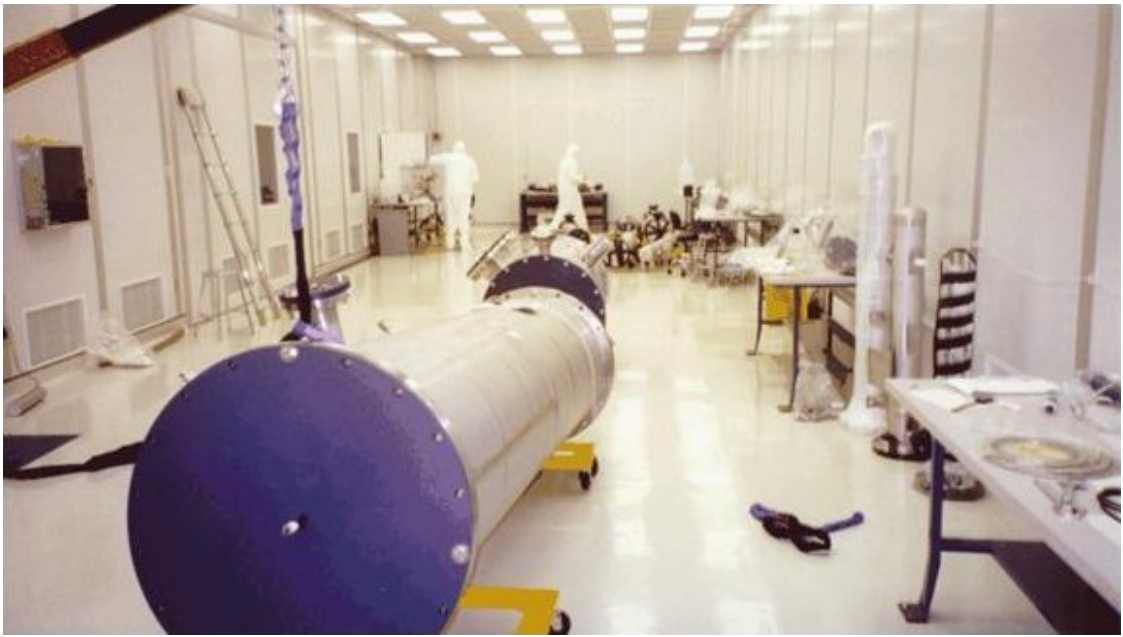

Figure 9. Assembly of the distillation column in the class 100 clean room at Princeton

After fabrication the equipment was disassembled and sent to AstroPak Corp. (Downey CA) for precision cleaning; cleanliness to level 50 Mil Std. 1246C was specified[34]. The stainless steel was also passivated to minimize the possible effects of corrosion. The exact procedures were proprietary but examples of cleaning and passivation procedures may be found in reference 28 and ASTM A967 and ASTM A380[32, 58]. The structured packing was cleaned by AstroPak in ultrasonic baths to the same specifications. Equipment was returned from AstroPak in sealed polyethylene packing. The internals of the distillation column, the water extraction column and the gas stripping column were also precision-cleaned to level 50 or better. The precision cleaned equipment was shipped back to Princeton where all the process equipment was reassembled in a 1000 sq. ft. class 100 clean room, and then sealed up and reconfigured into the skids. The system fit into two 10' x 10' x 40' skids. After cleaning and assembly, the skids were crated and shipped to Gran Sasso where they were hoisted and positioned in place. The skids were enclosed at Gran Sasso and fitted with a separate HEPA filtered air handling system to keep the interior of the skids at a Class 1000 clean room standard. After the skids were established as a clean room all the final connections were made, including the connections to the process lines in Hall C of LNGS.

Figure 9 shows the reassembly of the purification system inside the class 100 clean room at Princeton. Figure 10 shows one of the skids with the equipment remounted but sealed with polyethylene.

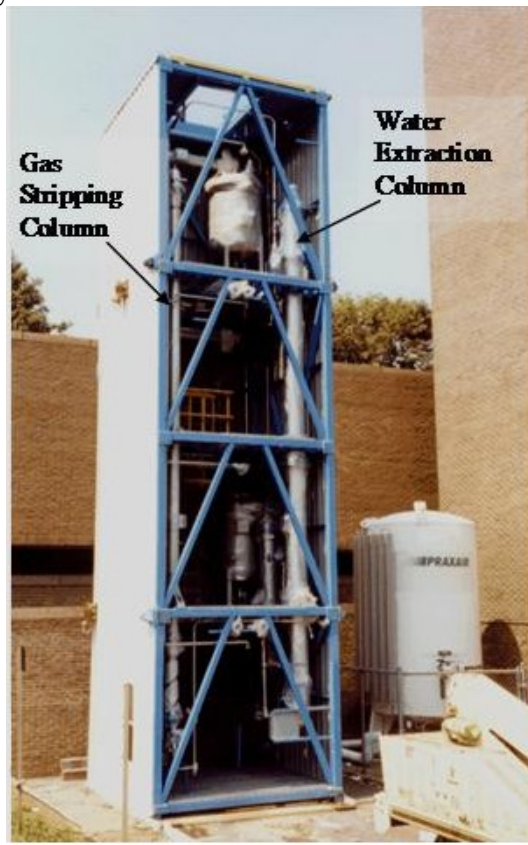

Figure 10. View of one of the two skids housing the purification system after reassembly at Princeton. The skids were shipped to Europe and brought into Hall C at Gran Sasso by truck.

*December 24, 2007*                                                                                          23

In addition to housing the process equipment the skids had to meet other safety requirements. The skid frame and equipment mounting were designed and certified for seismic safety. The skids were equipped with a fire safety system. Gas and liquid monitors were installed to sense hydrocarbon leaks. Catch basins were installed in the bottoms of the skids to contain any liquid spills. All system operations for purification have been reviewed with safety consultants and representatives from Princeton and Gran Sasso for Hazardous Operations (HazOps). The HazOps were used to identify potential problems during operation, and led to modifications for sensing and alarming the system. Any operation that could potentially result in a release of organic liquids or vapors to Hall C or the environment had to have two levels of safety alarms or be inherently safe (failure proof).

Auxiliary plants required by the purification system utilities were connected at Gran Sasso: the chilled water system for the condenser, the hot oil system for the reboiler in the distillation system and preheat for the gas stripping column, the nitrogen system for the instruments, pneumatic controls and pumps, the high purity nitrogen system for the gas stripping operations, the high purity water for water filling of the Borexino detector, the cleaning system for cleaning the system after the extended shutdown and a gas exhaust from vacuum pumps and nitrogen blankets.

In addition to the hardware a reliable control system is essential, considering the elevated temperatures that exist in the plant (in distillation mode), the flammability of the scintillator, and the enclosed environment in which the plant was located. The purification system is under the control of a master system for the Borexino experiment that provides for 24 hour/day operation, alarm notification and automated shutdown in case of problems. Details of the control system will be conveyed in a future publication.

## VII. COMMISSIONING AND OPERATION OF THE PURIFICATION SYSTEM

After installation at Gran Sasso the purification system goes through a series of tests and modifications to commission it for operation. Key for commissioning are cleaning and leak testing.

The purification system and all the equipment involved with the process (pumps, sensors, rupture discs, etc.) is cleaned in place and vacuum leak checked at the end of the cleaning campaign, in accordance with the leak-tightness requirements reported in section III.

Due to the complexity of the system and the large number of gasketed seals some connections are anticipated to come loose over time. In order to protect against such leaks two safety precautions are introduced.
1. Construction of aluminum enclosures with a continuous nitrogen purge over fittings and sensitive joints (sensors, rupture discs, flanges) that contact the scintillator;
2. Purging of all the double sealing gaskets with nitrogen.

After commissioning the purification system the nested nylon vessels of the Borexino detector were filled sequentially with nitrogen gas, DI water and finally scintillator and buffer. The vessels were maintained inflated during both water filling and filling with scintillator. Details of the liquid handling during filling operations will be reported in a future paper[53].

Borexino was filled with de-ionized (DI) water obtained from the Borexino high purity water system[59, 60]. The water was pumped to the Borexino scintillator purification system and stripped with nitrogen.

After nitrogen stripping the water was directed into the three detector volumes: 315.8 tons into the Inner Vessel (IV), 367.5 into the Inner Buffer (IB) and 643 into the Outer Buffer (OB). The filling was performed cyclically sending the water to the three zones. The flexible nylon vessels were all kept fully inflated with a nitrogen gas pressure above and water filling from below[23]. The water levels were kept the same in all three volumes to < 0.5 cm.

Borexino was filled with scintillator (PC + PPO 1.5 g/L) and with liquid buffer (PC + DMP 5g/L) by displacement of the water. The Purification System distilled and nitrogen stripped all 1167 tons of PC needed to fill the scintillator and buffer volumes of the detector. The buffer and scintillator volumes of the Borexino Detector were filled following a calculated table in order to replace 0.5 cm of water at a time with PC+PPO in the IV and PC+DMP in the Inner Buffer, Outer Buffer. The PPO and the DMP were added to the PC at the exit of the Purification System, through an automatic regulated mixing in line.

## VIII. EFFICACY OF THE BOREXINO PURIFICATION SYSTEM

Laboratory tests have been able to give guidance about the effectiveness of different purification methods[19, 20, 22, 35, 39, 61], but the laboratory tests cannot measure the level of impurities that are necessary for Borexino. The Borexino detector is the ultimate test of the purification system. The background levels of $^{14}$C, $^{40}$K, $^{238}$U, $^{232}$Th, $^{210}$Bi, $^{210}$Po, and $^{85}$Kr have all been determined after the initial filling of Borexino[14], and these background are sufficiently low to meet the requirements for measuring the $^7$Be solar neutrinos. These results are impressive and show that well known chemical purification methods can be employed to achieve extraordinary levels of radiopurity in liquid organic scintillators. Although the effectiveness of each specific purification method to remove radioactive impurities has not been identified, we have demonstrated that a multi-pronged approach extrapolated from well known processes applied in the petrochemical industry has proved successful.

# ACKNOWLEDGEMENTS

The committed support of the National Science Foundation through grants (PHY-0077423) has made the development of this purification system possible. The I.N.F.N. also provided support to complete the installation in the underground laboratory at Gran Sasso. We also thank G.P. Bellini, G. Ranucci and M. Pallavicini for their support and oversight at LNGS. The Borexino Steering Committee were vital in organizing and scheduling the successful filling and operation of the Borexino Detector. We also wish to thank George Schlowsky, Tom Shaeffer, Stan Lam and Jim DeNoble of Koch Modular Process Systems for their assistance with the design and construction of the Purification System, as well as their guidance in refining operational parameters during start-up. Domenico Barone, Roberto Tartaglia and Matthias Laubenstein provided invaluable guidance and assistance with the implementation of the safety systems. Ted Lewis and Lazlo Varga helped with much of the specialized machining necessary for the system. Stefano Nisi and Luca Ioannucci provided assistance with laboratory tests at LNGS.

# REFERENCES

1. Bahcall JN. *Neutrino Astrophysics*. Cambridge, England: Cambridge University Press; 1989.
2. Bahcall JN, Gonzalez-Garcia MC, Pena-Garay C. Global analysis of solar neutrino oscillations including SNOCC measurement. *Journal Of High Energy Physics.* Aug 2001(8).
3. Davis R. Solar Neutrinos.2. Experimental. *Physical Review Letters.* 1964;12(11):303-305.
4. Davis R. Nobel lecture: A half-century with solar neutrinos. *Reviews Of Modern Physics.* Jul 2003;75(3):985-994.
5. Hampel W, Handt J, Heusser G, et al. GALLEX solar neutrino observations: results for GALLEX IV. *Physics Letters B.* Feb 4 1999;447(1-2):127-133.
6. Abdurashitov JN, Bowles TJ, Cherry ML, et al. Measurement of the solar neutrino capture rate by SAGE and implications for neutrino oscillations in vacuum. *Physical Review Letters.* Dec 6 1999;83(23):4686-4689.
7. Hirata KS, Inoue K, Kajita T, et al. Results From 1000 Days Of Real-Time, Directional Solar-Neutrino Data. *Physical Review Letters.* Sep 10 1990;65(11):1297-1300.
8. Hirata KS, Inoue K, Kajita T, et al. Search For Day-Night And Semiannual Variations In The Solar Neutrino Flux Observed In The Kamiokande-Ii Detector. *Physical Review Letters.* Jan 7 1991;66(1):9-12.
9. Hirata KS, Kajita T, Kifune K, et al. Observation Of 8b-Solar Neutrinos In The Kamiokande-Ii Detector. *Physical Review Letters.* Jul 3 1989;63(1):16-19.
10. Ahmad QR, Allen RC, Andersen TC, et al. Measurement of the rate of nu(e)+d -> p+p+e(-) interactions produced by B-8 solar neutrinos at the sudbury neutrino observatory. *Physical Review Letters.* Aug 13 2001;8707(7):071301(071306).
11. Ahmad QR, Allen RC, Andersen TC, et al. Measurement of day and night neutrino energy spectra at SNO and constraints on neutrino mixing parameters. *Physical Review Letters.* Jul 1 2002;89(1):011301(011306).

12. Ahmad QR, Allen RC, Andersen TC, et al. Measurement of day and night neutrino energy spectra at SNO and constraints on neutrino mixing parameters. *Physical Review Letters.* Jul 1 2002;89(1).
13. Fogli GL, Lisi E, Montanino D, Palazzo A. Model-dependent and -independent implications of the first Sudbury Neutrino Observatory results. *Physical Review D.* Nov 1 2001;6409(9).
14. Borexino_Collaboration, Arpesella C, Bellini G, et al. First real time detection of 7Be solar neutrinos by Borexino. *Physics Letters B.* 2007:PLB 24344.
15. Borexino_Collaboration, Alimonti G, Arpesella C, et al. A large-scale low-background liquid scintillation detector: the counting test facility at Gran Sasso. *Nuclear Instruments & Methods In Physics Research Section A-Accelerators Spectrometers Detectors And Associated Equipment.* Apr 11 1998;406(3):411-426.
16. Borexino_Collaboration, Alimonti G, Anghloher G, et al. Ultra-low background measurements in a large volume underground detector. *Astroparticle Physics.* Feb 1998;8(3):141-157.
17. Borexino_Collaboration, Alimonti G, Arpesella C, et al. Science and technology of Borexino: a real-time detector for low energy solar neutrinos. *Astroparticle Physics.* Jan 2002;16(3):205-234.
18. Benziger JB. A scintillator purification system for a large scale solar neutrino experiment. *Nuclear Physics B-Proceedings Supplements.* Aug 1999;78:105-110.
19. Benziger JB, Johnson M, Calaprice FP, et al. A scintillator purification system for a large scale solar neutrino experiment. *Nuclear Instruments & Methods In Physics Research Section A-Accelerators Spectrometers Detectors And Associated Equipment.* Nov 11 1998;417(2-3):278-296.
20. Back HO, Balata M, de Bari A, et al. Phenylxylylethane (PXE): A high-density, high-flash point organic liquid scintillator for low energy neutrino experiments. *Nuclear Instruments and Methods in Physics Research Section A: Accelerators, Spectrometers, Detectors and Associated Equipment.* 2007;In Press, doi:10.1016/j.nima.2007.10.045
21. Buck C. *Development of metal loaded liquid scintillator for future detectors to investigate neutrino properties*. Heidelberg: Physics, Univ. Heidelberg; 2004.
22. Niedermeier L, Grieb C, Oberauer L, Korschinek G, von Feilitzsch F. Experimental scintillator purification tests with silica gel chromatography. *Nuclear Instruments and Methods in Physics Research Section A: Accelerators, Spectrometers, Detectors and Associated Equipment.* 2006;568(2):915.
23. Benziger J, L. Cadonati FC, E. de Haas, R. Fernholz, R. Ford, C. Galbiati, A. Goretti, E. Harding, An. Ianni, S. Kidner, M. Leung, F. Loeser, K. McCarty, A. Nelson, R. Parsells, A. Pocar, T. Shutt, A. Sonnenschein, R. B. Vogelaar. The Nylon Scintillator Containment Vessels for the Borexino Solar Neutrino Experiment. *Nuclear Instruments & Methods In Physics Research Section A-Accelerators Spectrometers Detectors And Associated Equipment.* 2007;582:509-534.
24. Ianni A, Lombardi P, Ranucci G, Smirnov OJ. The measurements of 2200 ETL9351 type photomultipliers for the Borexino experiment with the photomultiplier testing facility at LNGS. *Nuclear Instruments & Methods In*

*Physics Research Section A-Accelerators Spectrometers Detectors And Associated Equipment.* Feb 1 2005;537(3):683-697.
**25.** Brigatti A, Ianni A, Lombardi P, Ranucci G, Smirnov OJ. The photomultiplier tube testing facility for the Borexino experiment at LNGS. *Nuclear Instruments & Methods In Physics Research Section A-Accelerators Spectrometers Detectors And Associated Equipment.* Feb 1 2005;537(3):521-536.
**26.** Borexino_Collaboration, Arpesella C, Back HO, et al. Measurements of extremely low radioactivity levels in BOREXINO. *Astroparticle Physics.* Aug 2002;18(1):1-25.
**27.** Koshiba M. Observational Neutrino Astrophysics. *Physics Reports-Review Section Of Physics Letters.* Nov 1992;220(5-6):229-381.
**28.** Brarloutaud R. Status of the search for matter stability. *Nuclear Physics B (Proceedings Supplement).* 1992;28A:437-446.
**29.** Suzuki Y. Kamiokande Solar-Neutrino Results. *Nuclear Physics B.* Jan 1995;Suppl. 38:54-59.
**30.** Abundance of elements in Earth's crust. http://en.wikipedia.org/wiki/Abundance_of_elements_in_Earth's_crust.
**31.** Kanegsberg B, Kanegsberg E. *Handbook for Critical Cleaning*. Boca Raton, FL: CRC; 2001.
**32.** ASTM_International. Standard practice for cleaning, descaling, and passivation of stainless steel parts, equipment, and systems. Vol ASTM A380-99(2005): ASTM Internations; 2006:12.
**33.** Walters CT, Dulaney JL, Campbell BE. Advanced Technology Cleaning Methods for High-Precision Cleaning of Guidance System Components. In: Force A, ed. Vol Contract No. F04606-89-D-0034/DO Q807; 1993.
**34.** US_Army. Product cleanliness levels and contamination control. Mil Std. 1246C. In: US_Army, ed; 1994:26.
**35.** Leung M. Surface Contaminatin from Radon Progeny. *Topical Workshop on Low Radioactivity Techniques:LRT2004.* 2004;785:184-190.
**36.** Pocar A. Low background techniques for the Borexino nylon vessels. *Topical workshop on low radioactivity techniques:LRT 2004.* 2004;785:153-162.
**37.** Ranucci G, Goretti A, Lombardi P. Pulse-shape discrimination of liquid scintillators. *Nuclear Instruments & Methods In Physics Research Section A-Accelerators Spectrometers Detectors And Associated Equipment.* Aug 1 1998;412(2-3):374-386.
**38.** Ranucci G, Ullucci P, Bonetti S, Manno I, Meroni E, Preda A. Scintillation Decay Time And Pulse-Shape Discrimination Of Binary Organic Liquid Scintillators For The Berexino Detector. *Nuclear Instruments & Methods In Physics Research Section A-Accelerators Spectrometers Detectors And Associated Equipment.* Oct 15 1994;350(1-2):338-350.
**39.** Leung M. *The Borexino Solar Neutrino Experiment: Scintillator Purification and Surface Contamination*. Princeton: Physics, Princeton University; 2006.
**40.** Cimbak S, Cechova A, Grgula M, Povinec P, Sivo A. Anthropogenic radionuclides 3H, 14C, 85Kr, and 133Xe in the atmosphere around nuclear power reactors. *Nuclear Instruments & Methods In Physics Research Section B: Beam Interactions with Materials and Atoms.* 1986;17(5-6):560-563.

**41.** Loosli HH. A dating method with 39Ar. *Earth and Planetary Science Letters.* 1983;63:51-62.
**42.** Simgen H, Zuzel G. Ultrapure Gases - From the Production Plant to the Laboratory. *TOPICAL WORKSHOP ON LOW RADIOACTIVITY TECHNIQUES: LRT 2006.* 2006;897:45-50.
**43.** Perry RH, Green DW, Maloney JO, eds. *Perry's chemical engineers' handbook.* 7th ed. New York: McGraw-Hill; 1997.
**44.** McCabe WL, Smith JC, Harriott P. *Unit operations of chemical engineering.* 5th ed. New York: McGraw-Hill; 1992.
**45.** Borexino_Collaboration, Alimonti G, Angloher G, et al. Measurement of the C-14 abundance in a low-background liquid scintillator. *Physics Letters B.* Mar 12 1998;422(1-4):349-358.
**46.** Masetti F, Elisei F, Mazzucato U. Optical study of a large-scale liquid-scintillator detector. *Journal Of Luminescence.* Apr 1996;68(1):15-25.
**47.** Benziger J, Calaprice F, Johnson M, Shutt T. *Environmental Effect on the Optical Properties of Pseudocumene* May 1998. Borexino Research Report.
**48.** Shaeffer T, Schlowsky G. Proposal for a Pufication System for the Borexino Solar Neutrino Experiment. In: communication p, ed; 1997.
**49.** Schlowsky G. *Predicted tray efficiencies of the Borexino distillation column at reduced flow rates* 2005.
**50.** Clever HL. *Solubility Data Series, Volume 4: Argon Gas Solubilities.* Vol 4. New York: Pergamon Press; 1980.
**51.** Clever HL. *Solubility Data Series, Volume 2: Krypton, Xenon and Radon- Gas Solubilities.* Vol 2. New York: Pergamon Press; 1979.
**52.** Schlowsky G. Stripping calculation Gran Sasso Module. In: communication p, ed; 2002.
**53.** Borexino_Collaboration, Arpesella C, Bellini G, et al. Commissioning, Scintillator Purification and Filling of the Borexino Solar Neutrino Detector. *in preparation.* 2008.
**54.** Obara T, Miura T, Sekimoto H. Fundamental study of polonium contamination by neutron irradiated lead-bismuth eutectic. *Journal of Nuclear Materials.* 2005;343:297-301.
**55.** Wallner G, Berner A, Irlweck K. Aerosols: unexpected disequilibrium phenomena between airborne radio activities of lead-210 and its progenies bismuth-210 and polonium-210. *Naturwissenschaften.* 2002;89:569-574.
**56.** Giorgi AL, Bowman MG. A probe-type alpha detector for the measurement of polonium solutions. In: Commission USAE, ed; 1953.
**57.** Sutherland J. Surface Finish Terminology. *Cyberman: A Manufacturing Education Experience* [http://www.mfg.mtu.edu/cyberman/quality/sfinish/terminology.html.
**58.** ASTM_International. Standard specification for chemical passivation treatments for stainless steel parts. Vol ASTM967-05; 2005:6.
**59.** Giammarchi M, Balata M, Scardaoni R. High-purity water in a nuclear physics experiment. *Ultrapure Water Journal.* 1995;13:95.
**60.** Balata M, Cadonati L, Laubenstein M, et al. The water purification system for the low background counting test facility of the Borexino experiment at Gran Sasso.

*Nuclear Instruments & Methods In Physics Research Section A-Accelerators Spectrometers Detectors And Associated Equipment.* 1996;370:605.
**61.** Vogelaar R, Benziger J, Calaprice F, Darnton N. Removal of cosmogenic 7Be from scintillator. *Nuclear Instruments & Methods In Physics Research Section A-Accelerators Spectrometers Detectors And Associated Equipment.* 1996;372:59-66.